\documentclass[times, twoside]{MBpreprint}

\usepackage{polski}
\usepackage{blindtext}
\usepackage{float}
\usepackage{graphicx}
\usepackage{adjustbox}
\usepackage{graphicx}

\leadauthor{Borysiewicz} 

\begin{document}
\title{Mg nanostructures with controlled dominant c-plane or m-plane facets by DC magnetron sputter deposition} 
\shorttitle{Mg nanostructures with controlled dominant c-plane or m-plane facets}

% Use letters for affiliations, numbers to show equal authorship (if applicable) and to indicate the corresponding author

\author[1]{M.A. Borysiewicz $^{\text{\Letter},}$} \author[1]{P. Barańczyk} \author[1]{A. Wójcicka} \author[1,*]{J. Zawadzki} \author[1]{ M. Wzorek} \author[1]{R. Zybała}

\affil[1]{Łukasiewicz Research Network – Institute of Microelectronics and Photonics, Al. Lotników 32/46, 02-668 Warsaw, Poland}
\affil[*]{At time of manuscript submission no longer with the Łukasiewicz Research Network – Institute of Microelectronics and Photonics}

\maketitle

%TC:break Abstract
%the command above serves to have a word count for the abstract
\begin{abstract}
Magnesium nanostructures find increased use in applications for hydrogen storage, catalysis, waste treatment, and heat storage to name a few. Currently, most nanoparticles are made using a chemical synthesis approach, necessitating the use of organic solvents and yielding material covered in ligands. To apply these nanoparticles, one has to use them in paints or slurries for coating of surfaces, which again produces waste. In this communication we explore the possibilities of making magnesium nanostructures by a physical technique of magnetron sputtering and to control their crystallographic properties, i.e. the type of the dominating crystalline faces building up the external surface of the particle. We show that by applying different process parameters, it is possible to obtain dominating c-plane, mixed or dominating m-plane nanostructures. Since the surface-related adsorption processes are strongly related to the type of the crystalline plane, this report presents a clean, waste-free and large-scale approach to develop tailored nanostructured Mg coatings. 
\end {abstract}
%TC:break main
%the command above serves to have a word count for the abstract

\vspace{6pt}
\begin{corrauthor}
\color{color1}michal.borysiewicz@imif.lukasiewicz.gov.pl
\end{corrauthor}

%{\color{blue} jako komentarze w tekście}

\section*{Introduction}
Magnesium is a very light and strong Earth-abundant metal. Its mechanical properties make it useful for structural lightweight alloys for automotive and aerospace industries. It self-passivates with a thin impermeable oxide layer making it stable in air up to very high temperatures, although requires additional corrosion protection in harsh conditions. It ignites at 473 °C and burns with a very bright white light at a very high temperature of 2200 °C, making it applicable as an ignition agent for high ignition temperature mixtures such as thermite as well as the light source in flares and fireworks. It is studied as a material for temporary implants due to its biocompatibility and ability to dissolve by corrosion in vivo \cite{Kamrani2019}. These applications require strong and dense magnesium parts and coatings. However, the advantages of porous or nanostructured Mg are currently also being explored in other sectors. The material chemical properties and a highly developed surface enabling efficient interaction with the external atmosphere, electromagnetic radiation or chemical particles make nanostructured magnesium promising e.g. for: hydrogen storage in nano-magnesium hydrides \cite{Yartys2019}, \cite{Liu2014}, solar heat absorbers \cite{Hopper2022}, plasmonic-active media \cite{Ringe2020} catalysis \cite{Wang2023}, \cite{Santulli2022} or as precursor to oxidation into magnesium oxide nanostructures which can be applied in waste degradation, detection and degradation of pesticides, or antibacterial and antimicrobial applications in medicine, agriculture and food industries among others \cite{Fernandes2020, Suvarna2022, Chinthala2021, Thakur2022, Hornak2021}. 
The most common Mg nanostructures are nanoparticles, which can be made using wet chemical  route \cite{Ringe2020}, \cite{Lomonosov2023}, laser ablation into water \cite{Nyabadza2021}, or biosynthesis \cite{Rathore2015}, \cite{Sharma2023}, to name a few. As an outcome, material in the form of nano-powders is produced. To coat any surfaces areas with the material, one needs to add binder or other sort of adhesive and then develop the means of efficient and uniform coating. While these approaches give a large degree of material size control, and enable producing paints and inks for coating various organic substrates (e.g. for treating foliage with antimicrobial solutions) they require a significant amount of solvents throughout not only the synthesis process but also coating preparation, producing environmentally concerning waste. \\
For some of the applications where rigid surfaces need to be coated with Mg nanostructures, e.g. for waste treatment, hydrogen storage, catalysis, etc. vacuum-based techniques could be taken into account. Although not popular in the development of nanostructures, vacuum techniques such as sputtering have been shown to deliver nanostructures of materials such as Zn \cite{Masyk2016} or Mg \cite{Ham2014}. The application of vacuum processing for the fabrication of large-scale nanostructured porous films while requiring more costly equipment, yields materials which are inherently attached to the substrate in a binder-free way and utilize no solvents in the process making in a green alternative. The surface of the nanostructures is also more clean due to no ligand presence than in the case of solvent-based approaches. This makes the vacuum-based approach considerable for large-scale applications. As far as porous sputter-deposited Mg has been reported, the view was rather that the porosity was detrimental due to the perceived applications of Mg films where high film uniformity and density is required, which is why high ionization techniques (HIPIMS) have been successfully applied to get dense Mg films \cite{Moens2019}. 
In this communication we wish to explore the possibilities of treating the porosity and nanostructure of the sputter-deposited Mg films as positive and see how the yet unstudied process properties influence the morphology and structure of the nanostructures and in particular show that it is possible to tailor the dominant crystalline plane which is exposed to the environment, which can be relevant for all absorption applications.

\section*{Experimental details}
Opposed to the literature, where Mg deposition only takes place in pure Ar gas, we use an Ar-O$_2$ mixture, akin to our experiments with obtaining porous Zn films with such a mixture \cite{Masyk2016}, which we will reference here. We also apply variable substrate temperature to see how it will influence the nanostructure of the films. The deposition reactor is a Surrey Nanosystems Gamma 1000C with four confocal magnetrons at a distance of 14 cm from the middle of the 6” substrate table. The substrates used were cleaned Si (111) and the substrates were rotated during deposition. We use one magnetron with a 99.99$\%$ pure Mg target 3” in diameter as the source. The magnetron is supplied with 80W DC power. The Ar and O$_2$ gases are 99.9999$\%$ pure. We explore different ratios of the Ar:O$_2$ flows, including 10:1, 15:1 and 10:2. The total sputtering pressure is controlled independently from the gas flows by means of a Baratron-controlled throttle valve in front of the cryo-pump. We assessed the porosity of the films for total pressures between 0.2 Pa and 3.2 Pa and chose 0.4 Pa as optimal for all experiments. Finally, we show that by applying increased substrate temperature during growth it is possible to grow films of nanowires of Mg. The base vacuum in all cases before film growth was of the order of 10$^{-5}$ Pa. 
The structure of the films is showed by means of X-ray diffraction (XRD) in a Bragg-Brentano geometry using a Panalytical Empyrean with a Cu anode. Scanning Electron Microscope (SEM) images of the cross-sections and top surface of the films are taken using a Zeiss Gemini and Transmission Electron Microscope (TEM) images of selected crystallites with their surface oxide passivation are done using a JEOL JEM 2100. 
To see if any insights could be taken from the differences between the well understood Zn and new Mg growth, Monte Carlo SIMTRA \cite{Strijckmans2014} simulations were performed to assess the number of particles and their mean energy for the two systems under the same deposition conditions. 
\begin{table*}%[H]
\centering
\caption{Description of deposition process parameters for samples in batch I.}
\label{T1}
\begin{tabular}{|l|l|l|l|l|l|l|l|}
\hline
\textbf{Sample} & \textbf{Ar (sccm)} & \textbf{O$_2$ (sccm)} & \textbf{Flow ratio} & \textbf{p (Pa)} & \textbf{P (W)} & \textbf{U (V)} & \textbf{Avg. size (nm)} \\ \hline
I.A             & 3                 & 0.6               & 10:2                & 0.4            & 80            & 236            & 85                      \\ \hline
I.B             & 6                 & 0.6               & 10:1                & 0.4            & 80            & 224            & 248                     \\ \hline
I.C             & 9                 & 0.6               & 15:1                & 0.4            & 80            & 220            & 256                     \\ \hline
\end{tabular}%
\end{table*}

\section*{Results and discussion}
\begin{figure}%[H]
    \centering
    \includegraphics[scale = 0.8]{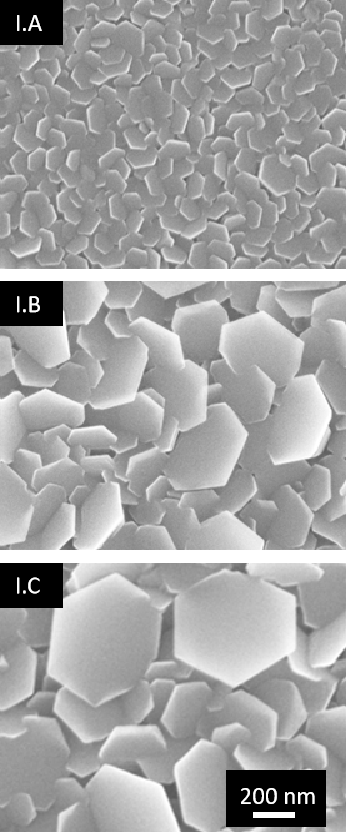}
    \caption{SEM plan-view images for the Mg films deposited in batch I. For identification please refer to Table \ref{T1}.}
    \label{fig1}
\end{figure}

The initial conditions for deposition were similar to those upon which we observed porous Zn growth, i.e. 6 sccm : 0.6 sccm Ar : O$_2$ flow, 80W DC power and a total gas pressure of 0.4 Pa \cite{Borysiewicz2012}. To see how the argon to oxygen ratio influences the morphology, we chose 0.6 sccm O$_2$ as the set oxygen flow, and we changed the argon flow rates from 3 sccm (10:2 flow ratio), through 6 sccm (10:1 flow ratio) to 9 sccm (15:1 flow ratio). For a description of the samples in this batch, denoted as I.A, I.B and I.C, please see Table \ref{T1}. Contrary to previous reports where the role of O$_2$ was not studied, we see that the oxygen presence during sputtering influences the sizes of the nanofeatures. The hexagonal-like platelets are similar to the ones reported for deposition in pure Ar. Mg has a hexagonal crystal lattice and the visible six-sided platelet faces are the (0001) crystalline c-planes. Not much is seen in plan-view of the side (10-10) crystalline m-planes. The c-face has the lowest surface energy of the Mg crystal faces \cite{Du2017}, therefore the system wants to have the largest c-faces possible and grows in the shape where the hexagonal c-face is big and the sides are very short. The platelets are randomly oriented and they cross one another. The only visible morphology differences are in the platelet size: for 10:2 flow ratio, i.e. a rather large oxygen percentage, the sizes are smallest with the mean edge-to-edge width equal to 85 nm. Decreasing the oxygen percentage by raising the argon flow, moving to 10:1 and then 15:1 argon to oxygen flow ratios, the platelet widths become 248 nm and 256 nm indicating a more rapid growth of the structures (see Fig.\ref{fig1}). At the same time, we see a decrease in target voltage from 236 V to 224 V and 220V indicating lower target surface poisoning, confirming that the deposition speed controls the size of the Mg nanostructures, as previously suggested for an oxygen-free system \cite{Moens2019}. 

This behavior is different to what we observed for the Zn nanostructure formation, where both the target-based and growth front-based phenomena influenced the formation and morphology of the nanostructures. There, oxygen presence in the atmosphere acted as growth inhibitors on the nanocrystals, promoting many nucleation centers and growth of smaller grains. In the case of Mg, this is not seen.

\begin{figure}[H]
    \centering
    \includegraphics[scale = 0.27]{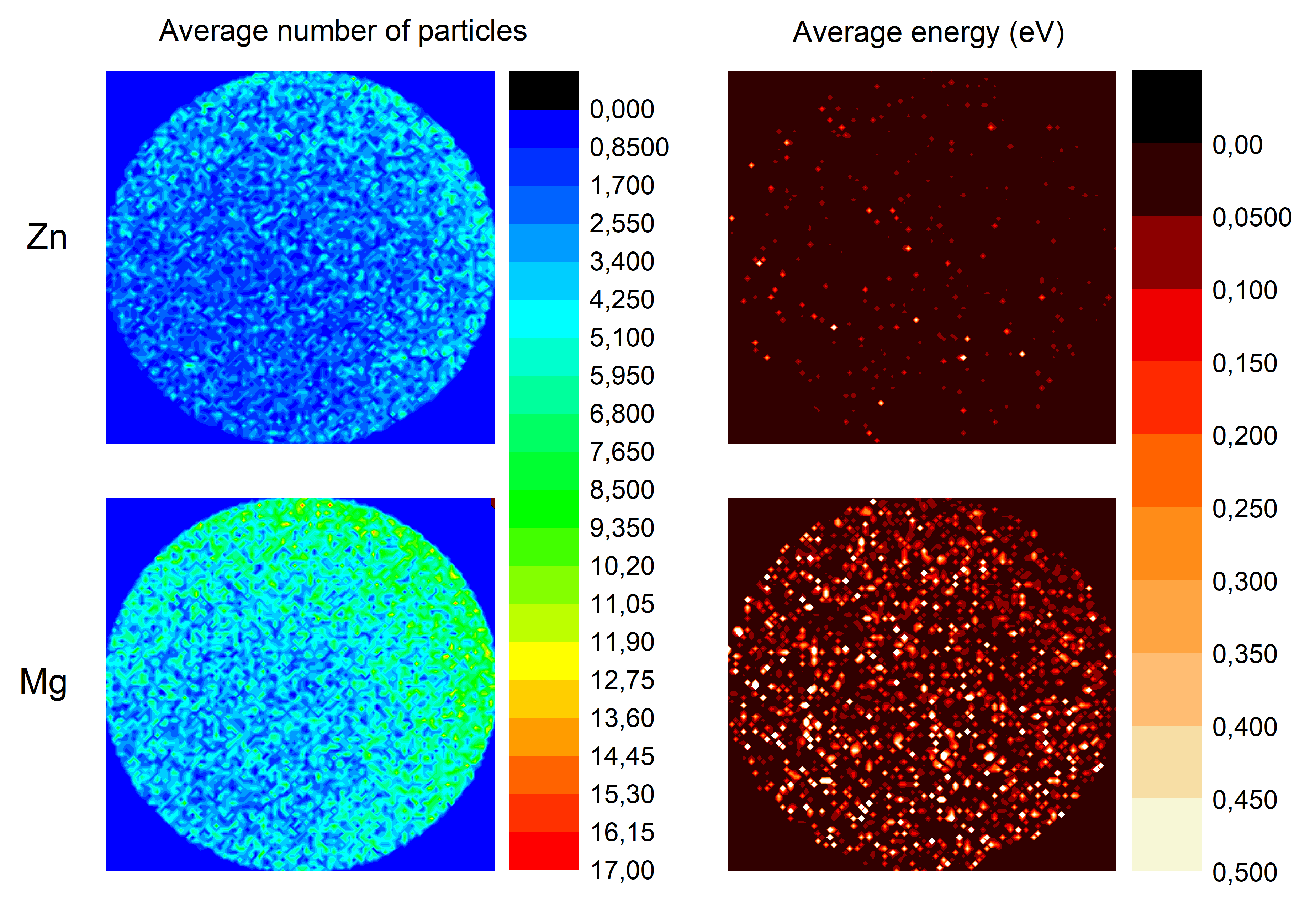}
    \caption{SIMTRA-simulated average number of particles and average energy of particles for Mg and Zn sources at 22.5° to the right. The color scales are common for each Zn and Mg image pairs in the two data categories.}
    \label{fig2}
\end{figure}

To understand better the target-based behavior of the material, we performed simulations of the average number of particles and average energies for both Zn and Mg ejected from a 3” target onto a 6” receiving plane being the sample table in the deposition chamber. The dimensions are true to the real sputtering chamber dimensions used and the magnetron is at 22.5° off to the right side, aiming at the middle of the receiving plane (see Fig.\ref{fig2}). We see that for Mg both the flux and the energy of the particles are significantly higher than in the case of Zn, meaning that the deposition rate is much higher and the incoming particles are much more energetic, yielding potentially larger crystallites due to the additional energy for adatom diffusion on the growth front. Even taking into account that magnesium is more reactive than zinc in the metal reactivity series, the difference in the number and energy of particles seems to be large enough that the oxygen does not have the time needed to actually react with or trap and force nucleation of new crystallites. 

\begin{figure}%[H]
    \centering
    \includegraphics[scale = 0.8]{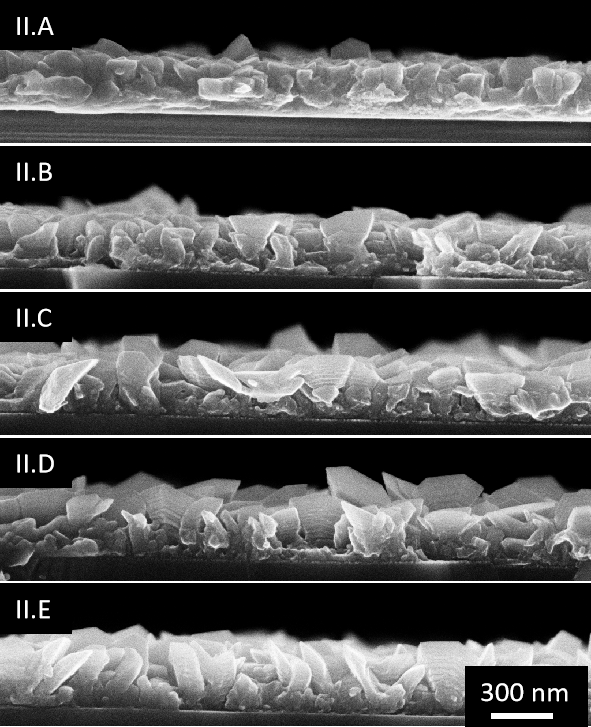}
    \caption{SEM cross-section images for the Mg films in batch II. For identification please refer to Table \ref{T2}. }
    \label{fig3}
\end{figure}

\begin{figure*}
    \centering
    \includegraphics[scale = 0.6]{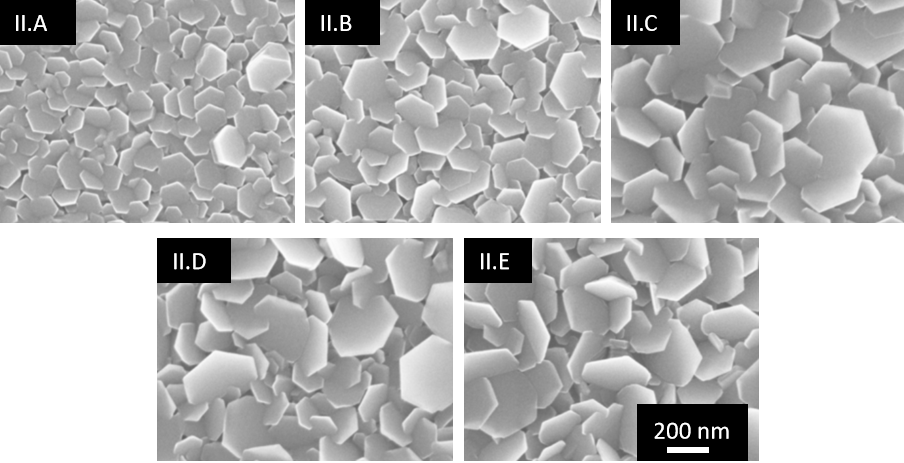}
    \caption{SEM plan-view images for the Mg films in batch II. For identification please refer to Table \ref{T2}. }
    \label{fig4}
\end{figure*}

To further explore the role of oxygen content and flow values we deposited samples at a constant argon to oxygen flow ratio of 15:1 and changed the value of the actual gas flow rates, changing the amount of the available gas in the chamber and its dwell time - for larger flows, the gas molecules spend less time in the chamber. The sample deposition parameters for this batch, number II, are summarized in Table \ref{T2}. The cross-section and plan-view SEM images of the films are presented in Fig. \ref{fig3} and Fig. \ref{fig4}, respectively. From the images it can be seen that the rate of crystallite growth and the average lateral size of the crystallites increase with the increase in the gas flow values. Also, for the highest flow values used, in sample II.E, the nanostructures grow in a more perpendicular direction to the substrate than for the lowest flow values. 

\begin{figure*}
    \centering
    \includegraphics[scale = 1.0]{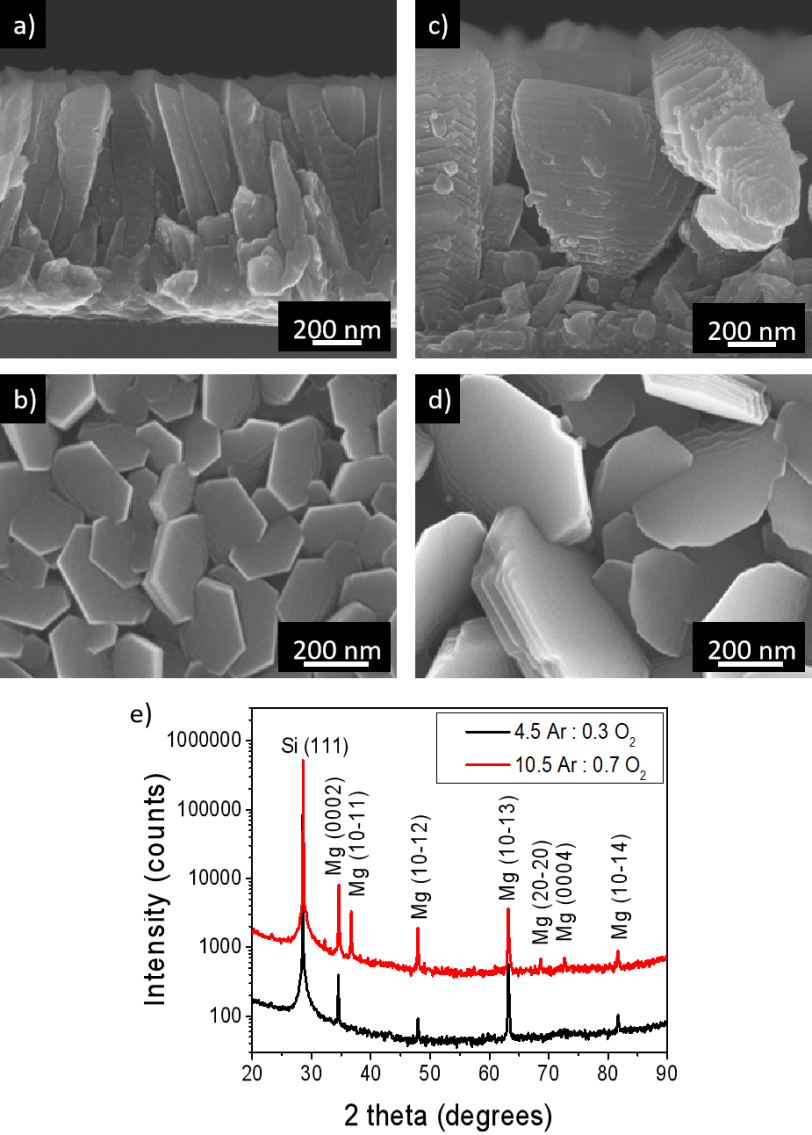}
    \caption{SEM plan-view and cross-section images for the Mg films deposited at 4.5 accm Ar : 0.3 sccm O$_2$ (a, b) and 10.5 sccm Ar : 0.7 sccm O$_2$ for 4 times longer time (c, d).  (e) XRD pattern of the deposited films.}
    \label{fig5}
\end{figure*}

The platelet width is similar to the one in sample II.C and II.D, and the film thicknesses are also similar but the orientation of the whole grain changes. This difference is even more pronounced when the growth is continued for an hour - four times more than for batch II - see Fig.\ref{fig5}: for sample of the type II.A, smaller crystallite faces are evident, and their orientation is more parallel to the substrate surface than in the case of type II.E sample, where side-step features are visible. The XRD patterns of the materials confirm the changes with the pattern for lower flow values presenting a material more oriented with just the main peaks visible (0002) and (10-13), whereas the material deposited at higher flow values has more peaks in the diffraction pattern. This means that by changing the flow values it is possible to tailor the crystalline character of the films, which might be beneficial for applications requiring selective absorption processes which can take place at different planes with different efficiency. 
\begin{table*}
\centering
\caption{Description of deposition process parameters for samples in batch II.}
\label{T2}
%\resizebox{\columnwidth}{!}{%
\begin{tabular}{|l|l|l|l|l|l|l|}
\hline
Sample & Ar (sccm) & O$_2$ (sccm) & P (W) & U (V) & Avg. size (nm) & Thickness (nm) \\ \hline
II.A   & 4.5      & 0.3      & 80   & 224   & 122            & 288            \\ \hline
II.B   & 6        & 0.4      & 80   & 224   & 185            & 330            \\ \hline
II.C   & 7.5      & 0.5      & 80   & 220   & 267            & 343            \\ \hline
II.D   & 9        & 0.6      & 80   & 220   & 268            & 357            \\ \hline
II.E   & 10.5     & 0.7      & 80   & 220   & 264            & 358            \\ \hline
\end{tabular}%
%}
\end{table*}

\begin{figure}%[h]
    \centering
    \includegraphics[scale = 0.6]{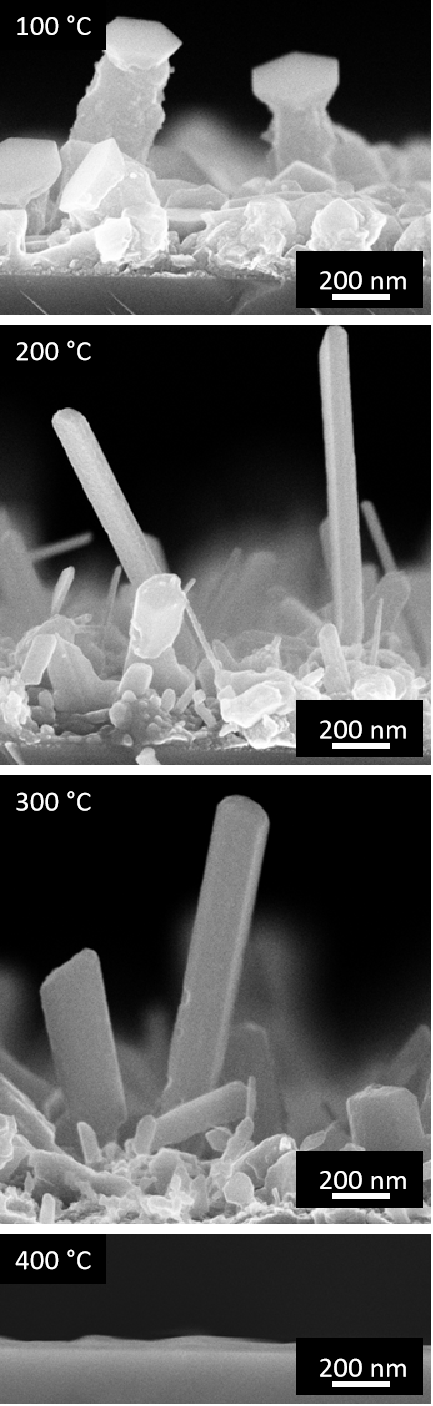}
    \caption{SEM cross-section images of the samples deposited at 7.5 accm Ar : 0.5 sccm O$_2$ at the temperatures of 100 °C, 200 °C, 300 °C, 400 °C.}
    \label{fig6}
\end{figure}

\begin{figure}%[H]
    \centering
    \includegraphics[scale = 0.6]{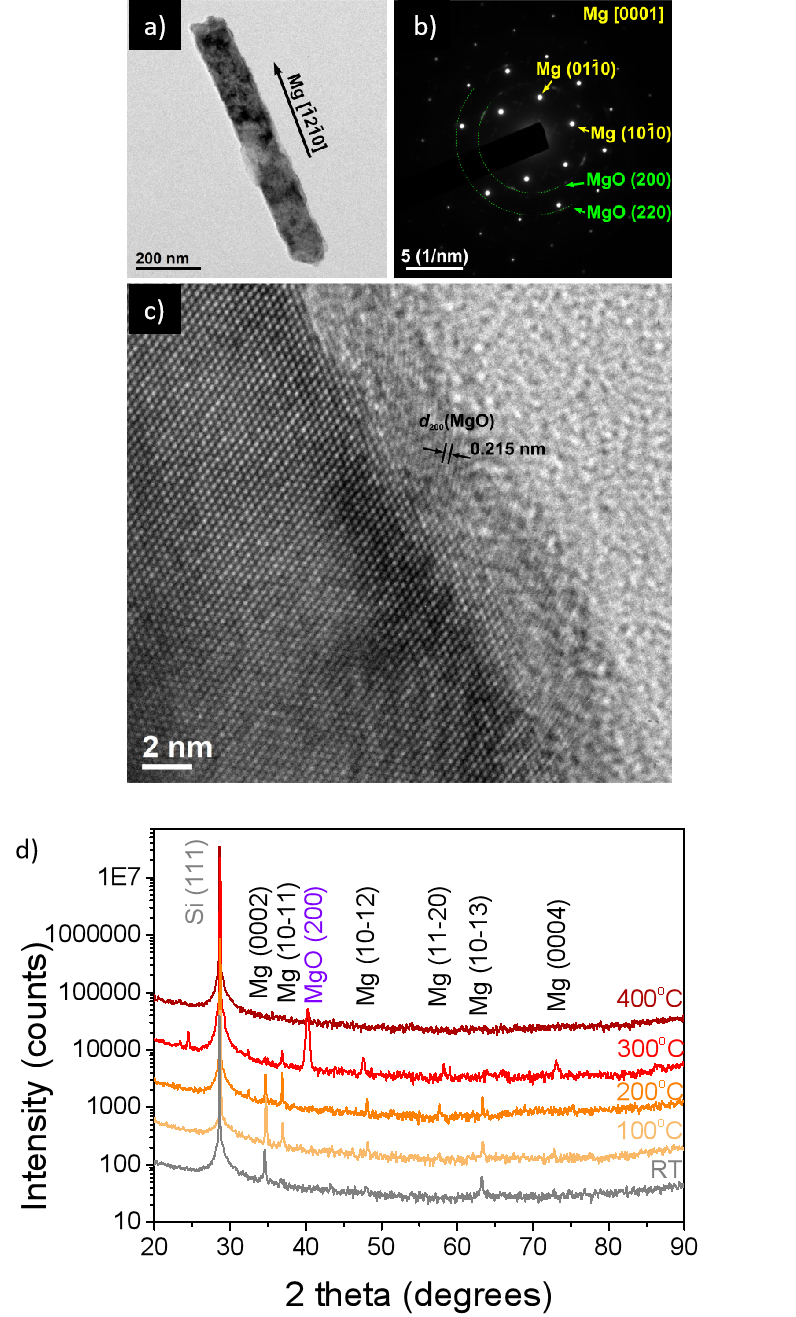}
    \caption{TEM image of a single nanowire deposited at 200 °C (a), selected area electron diffraction pattern (b) and a high-resolution close-up of the surface area of the nanowire, showing a thin layer of a native oxide (c). (d) XRD pattern of the films deposited at elevated temperatures and without heating (RT).}
    \label{fig7}
\end{figure}

The mechanism of these changes is not controlled by the target-based phenomena, in particular target face poisoning, since the target voltages do not change significantly, as was the case for batch I. This might be related to the different nucleation of the initial stages of crystallization due to the lower gas dwell times interacting with the growth front as discussed by Moens et al. \cite{Moens2019}.\\
The final process parameter studied, which influenced the nanostructure growth significantly, and enabled to achieve nanowire growth, was the substrate temperature. We deposited a third batch of samples at the conditions similar as for sample II.C, i.e. 7.5 sccm Ar : 0.5 sccm O$_2$ at the temperatures of 100 °C, 200 °C, 300 °C, 400 °C and a reference sample without heating. We did not pursue temperatures higher than 400 °C since the autoignition temperature of Mg in air is around 473 °C \cite{IPCSINCHEM2000} and the gas mixture inside the deposition chamber contained oxygen. An autoignition in the deposition chamber could be dangerous for the equipment. For all samples the target voltage was constant at 216 V, indicating no changes in the sputtering process. Therefore the only changes in the resulting nanostructures are results of effects happening at the growth front. \\
Analyzing the changes in morphology when increasing the substrate temperature we can see that the crystallites start to elongate, at lower temperatures still exhibiting the hexagonal facets at the end, and at the higher ones - the features look like nanowires. The nanowire density gets smaller as the temperature increases and finally, at 400 °C the film looks featureless (see Fig.\ref{fig6}) an there seems to be no film. An explanation of this behavior can be found in analyzing the surface energies of Mg. With the increase in temperature additional energy is added to the system. This increases the energy threshold for the growing facets, promoting the growth of surfaces with higher surface energy. In the case of Mg, the surface energy of the (10-10) planes is 0.734 J·m$^{-2}$, whereas fot the (0001) planes it is 0.703 J·m$^{-2}$ \cite{Du2017}. In the increased temperatures the (0001) planes are thus less available and the (10-10) planes start to dominate, wchich yields nanowire formation, as this form has a big m-plane surface area. High-resolution transmission electron microscope images and selective area electron diffraction patterns for a selected nanowire show that the material is mostly metallic single crystalline Mg, with a 3 nm surface coverage of MgO, which most probably is due to the ambient oxidation after removal of the material from the vacuum chamber (see Fig.\ref{fig7}). For the samples deposited without heating and those deposited at 100 °C and 200 °C, no MgO inclusions or otherwise oxygen-rich sections were identified in the bulk of the nanowires meaning that the oxygen does not react with the material during growth, even at temperatures. \\
However, when reaching 300 °C the nanowires convert to MgO, as also evidenced by the X-ray diffraction patterns (see Fig.\ref{fig7}.d). The nanowire films grow with oxygen participation at the growth front and that MgO is formed. A small presence of Mg-related peaks shows that the material is not 100\% pure MgO. 
The lack of films at 400 °C is related to lower Mg sublimation temperature at the high vacuum conditions in the sputtering chamber, than it would be at ambient pressure - lower than its ignition point. The material does not want to stick to the substrate and no film is observed.

\section*{Conclusions}
By studying the DC magnetron sputtering deposition of magnesium we were able to describe the process parameters influencing mostly the nanostructure of the deposited films. We show that the gas dwell time plays an important role in controlling the size and orientation of the crystallites which generally have the shape of c-plane platelets rather parallel to the surface of the substrate. With increased flow values the c-planes become more perpendicular to the substrate surface, exposing large areas of m-plane steps. On the other hand, the oxygen content in the sputtering atmosphere does not seem to influence the results significantly other than through potential target poisoning, leading to lower growth rates and smaller resulting crystallites. The most striking influence on the film morphology was through substrate heating during deposition. Substrate temperatures of 200-300 °C yield films of nanowires with different orientation to the substrate but with large m-plane side faces. 
Therefore, we show the way of growing large area coatings of Mg nanostructures with either c-planes, m-planes or a mix of c and m-planes exposed to the environment. In applications where the Mg nanostructures absorb different volatile species, such as in sensors or hydrogen storage, such control of the crystalline properties may be key to achieve high performance. In particular, it was shown recently that nonpolar Mg surfaces exhibit significantly higher hydrogen uptake than the polar (0001) ones \cite{Dun2022}. The results presented in this report could thus help obtain large-scale hydrogen storage coatings with improved efficiency. 

\section*{Acknowledgments}
This research was funded in whole by the National Science Centre, Poland, under Grant number UMO-2020/39/D/ST5/01474. For the purpose of Open Access, the author has applied a CC-BY public copyright licence to any Author Accepted Manuscript (AAM) version arising from this submission.

\section*{Bibliography}
\bibliography{Porto-Mg.2}

\end{document}